\documentclass[aps,prl,twocolumn,amsmath,amssymb,floatfix]{revtex4}
\usepackage{epsfig}
\usepackage{bm}
\def\l{\langle\!\langle}
\def\r{\rangle\!\rangle}
\def\la{\langle}
\def\ra{\rangle}
\def\be{\begin{equation}}
\def\ee{\end{equation}}

\begin{document}
\title{Achieving the threshold regime with an over-screened Josephson junction}
\author{Eugene V. Sukhorukov$^1$ and Andrew N. Jordan$^2$}
\affiliation{$^1$\ D\'{e}partment de Physique Th\'{e}orique, Universit\'{e} de Gen\`{e}ve,
CH-1211 Gen\`{e}ve 4, Switzerland\\
$^2$\ Department of Physics and Astronomy, University of Rochester, Rochester, 
New York 14627, USA }
\date{\today}

\begin{abstract}
We demonstrate that by utilizing an over-screened Josephson junction as a noise detector it is possible to
achieve the threshold regime, whereby the tails of the fluctuating current distribution are measured.  This
situation is realized by placing the Josephson junction and mesoscopic conductor in an external circuit with
very low impedance.  In the underdamped limit, over-screening the junction inhibits the energy diffusion in
the junction, effectively creating a tunable activation barrier to the dissipative state.  As a result, the
activation rate is qualitatively different from the Arrhenius form.
\end{abstract}


\maketitle
In the course of scientific progress, it is desirable to use recent advances of physical understanding to
develop new experimental techniques that will in turn give impetus for further advances.  This description
is apt for the current state of electron counting statistics in mesoscopic physics, and the aim of the
current Letter. Recent experiments in electron counting statistics have measured the asymmetry of the
current distribution \cite{reulet1,Bomze}, and the effect of the measurement circuit
\cite{reulet1,reznikov}.  
%Further developments have taken several independent  
Subsequent researchers have taken several independent
approaches to exploring this
physics.  If the system is transferring electrons on sufficiently long time scales, it is possible to
directly count individual electrons \cite{Simon1}, look at various higher current cumulants \cite{Simon2},
and even examine the conceptually new area of conditional counting statistics \cite{conditional}.  It is
also possible to focus on the frequency dependence of the cumulants \cite{pilgram,reulet2}.  A further
exciting possibility is to measure the tails of the current fluctuation distribution via a threshold
detector \cite{Nazarov,ourPRB}.

A natural threshold detector is a Josephson junction (JJ): by measuring the rate of switching out of the
metastable supercurrent state into the running dissipative state, information about the statistical
properties of the noise driving the system may be extracted. This system has been the subject of recent
experiments \cite{Pekola,Pekola2,Saclay}.   However, it was noticed that a single JJ typically works near the
Gaussian point \cite{Nazarov} (a regime that has been well-studied in the past, see e.g. \cite{markusjj,dykman}),
so much so that the third current cumulant makes only a small correction to the escape rate 
\cite{Ankerhold,SJthreshold1,grabert}.  This is because the slow semiclassical dynamics of the JJ averages out the Markovian noise source and becomes effectively Gaussian under typical conditions.  The purpose of this Letter is to show that despite this difficulty, the {\it threshold regime} is realizable for a single underdamped JJ.  The threshold regime goes beyond the third
cumulant and realizes the full potential of the Josephson detector, where the escape to the running state of
the JJ is driven by the tails of the distribution, rather than by relatively small deviations from the
average.

\begin{figure}[tb]
\epsfxsize=7.1cm
\epsfbox{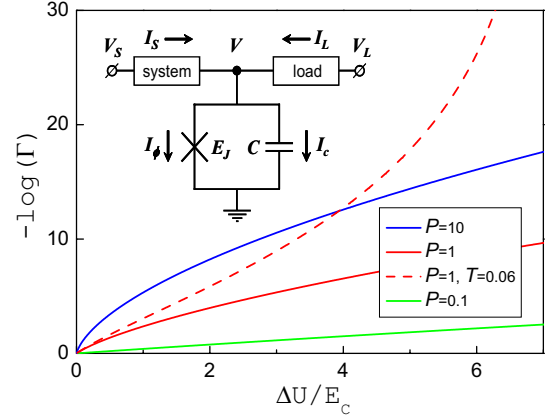}
%\epsfbox{fig1.eps}
\caption{(Color online) The negative logarithm of the escape rate is plotted versus dimensionless barrier height $\Delta U/E_C$ for various values of the threshold parameter ${\cal P}$.  The solid lines correpond to Poissonian noise, while the dashed line corresponds to a QPC with transparency ${\cal T} = 0.06$.
Inset: Simplified electrical circuit for the Josephson junction (JJ, marked with an X) threshold detector of the noisy mesoscopic system.   
}
\vspace{-5mm}
\label{circuit}
\end{figure}

Circuit effects are known to be important for electron counting statistics.  The measurement circuit can lead to cascade corrections \cite{nagaev,Beenakker} to higher current cumulants, masking the system contribution.  A similar effect has been predicted by the authors to occur in JJ detectors \cite{SJthreshold1}.  The central idea of this Letter is to have a very small load resistance, resulting in an over-screened junction, whilst still maintaining the underdamped state (see inset of Fig.\ \ref{circuit}). The advantage is that not only cascade corrections are suppressed but also the relative contribution of the third cumulant is enhanced compared to the Gaussian contribution  \cite{SJthreshold1}, assisting in the measurement of the third cumulant in recent experiments \cite{Pekola2,Saclay}.  We will demonstrate that this is the case not just for the third cumulant, but that all higher cumulants are also enhanced, leading to non-perturbative activation by rare events.

%We now briefly summarize our main results.  Considering the semi-classical JJ in the weak damping limit for a low load-impedance circuit, we find the escape rate for a general potential using the stochastic path integral formalism \cite{spi}. We show that for the case of Poissonian noise produced by a tunnel junction the activation is dominated by the rare current events, which leads to the breakdown of Arrhenius behavior for the escape rate.  We describe stabilization effects for both binomial shot noise and telegraph noise. The ranges of experimental parameters required to see these effects are also given. 

{\it Josephson detection circuit.}---
%Before proceeding to our results, we first review the physics of the Josephson junction circuit.
The circuit in the inset of Fig.\ \ref{circuit} shows the essential part of the detector comprised of the JJ with
Josephson energy $E_J$, and the capacitor, $C$. The fluctuations in the circuit originate from the
combination of the current $I_L$ through the macroscopic load resistor and by the mesoscopic system current
$I_S$, which is to be measured. According to Kirchhoff's law, the total current $I_S+I_L$ is equal to
the sum of the Josephson current $I_{\phi}=(E_J/\Phi_0)\sin\phi$ where $\Phi_0=\hbar/2e$, and the
displacement current $I_C=C\dot V$. This leads to the equation of motion for the superconducting phase
$\phi$,
\begin{equation}
C\Phi_0^2\ddot\phi+E_J\sin\phi=\Phi_0(I_S+I_L),
\label{motion}
\end{equation}
where we used the relation $V=\Phi_0\dot\phi$. 

We consider an ohmic system and load resistor, so that $\langle I_S\rangle=J_S-G_S V$ and $\langle
I_L\rangle= J_L-G_L V$, where $G_S$, $G_L$ are the system and load conductances, and the constant currents
$J_S=G_S V_S$, $J_L=G_L V_L$ are tunable parameters that will be shown to control the activation threshold.
Equations~(\ref{motion}) can be rewritten as a set of Hamiltonian-Langevin equations for the phase
variable $\phi$ and canonically conjugated momentum $p=\Phi_0 Q$ (where $Q=CV$ is the total charge on the
capacitor),
\begin{equation}
\dot\phi=p/m,\quad \dot p=-\partial U/\partial\phi + \Phi_0 \delta I,
\label{HLE-phase}
\end{equation}
with ``mass'' $m=\Phi_0^2C$, and where $\delta I = I_S-J_S + I_L - J_L$ is the dissipative part of the system and
load current.  Equations (\ref{HLE-phase}) describe the motion of a ``particle'' in the tilted periodic 
potential
\begin{equation}
U(\phi)=-E_J [\cos\phi + {\cal J} \phi],
\label{potential}
\end{equation}
stimulated by the dissipative part of the system and load current, where ${\cal J}=\Phi_0(J_S+J_L)/E_J$ is
the dimensionless total current bias.  Dissipation leads to relaxation of the JJ into one of its metastable
supercurrent states, where the phase is localized in one of the potential wells so that $\langle V\rangle=0$. 
In the dissipative state, the phase drifts along the bias which generates a non-zero voltage drop $V$.

{\it Weak damping threshold limit.}---
Here the phase oscillates with the plasma frequency 
$\omega_{\rm pl}=\Omega_J(1-{\cal J}^2)^{1/4}$, 
where $\Omega_J=\sqrt{E_J/m}$.  The energy relaxes slowly with 
rate $(G_S+G_L)/C<\omega_{\rm pl}$ 
to the local potential minimum. We further assume 
the separation of time scales, $\max\{eV_S, T\}>\hbar\omega_{\rm pl}$, 
so that the noise source $I_S$ is Markovian. According to our previous 
results \cite{SJthreshold1}, the escape rate $\Gamma$ (predicted to exponential accuracy) is
\begin{equation}
\log\Gamma=-\Phi_0^{-1}\int \lambda dE, 
\quad \langle{\cal H}(\lambda\dot\phi)\rangle_E=0.
\label{WDR-instanton}
\end{equation}
Here,  the function ${\cal H}(z)$, which we refer to as a Hamiltonian, generates the cumulants (irreducible moments) of $\delta I$ in Eq.~(\ref{HLE-phase}), so that $\partial_z^n {\cal H}|_{z=0} = \l \delta I^n \r$.  The function $\lambda(E)$ is the escape trajectory in the extended energy phase-space, or the ``instanton line''.  The notation $\la \ldots \ra_E = (1/{ T_{\rm p}}) \oint dt (\ldots)$ denotes time-averaging over a physical
trajectory in the $(q, p)$ phase-space at a certain energy $E$, with ${ T_{\rm p}}(E)$ being the period of the quasi-periodic motion at that energy.  See Ref.~\cite{SJthreshold1} for a detailed derivation.

We note that after replacing the node voltage $V$ with $\Phi_0 {\dot \phi}$, the first two coefficients in
the expansion of ${\cal H}$ are (i) $\la \delta I \ra = - (G_S + G_L) \Phi_0 {\dot \phi}$ and (ii) 
$\l\delta I^2 \r = 2G_L T + \l I_S^2 \r$. We now focus on the limit of small load resistance, $G_L \gg G_S$
so circuit back action can be neglected \cite{SJthreshold1}.  In this
case, the load controls coefficient (i).  However we observe that for a large system voltage, the
system shot noise will dominate the load resistor noise in coefficient 
(ii) if $eV_S > G_L T/G_S$.  The circuit load, being a macroscopic resistor, has no higher current 
cumulants, so the Hamiltonian takes the form 
$\la {\cal H} (\lambda \dot\phi) \ra_E = -G_L \Phi_0 \lambda \la {\dot \phi}^2\ra_E +
J_S \la F(\lambda{\dot \phi})\ra_E$, 
where $F$ generates the generalized Fano factors of the system,
\be
F(z) = \sum_{n=2}^\infty \frac{z^n}{n!} \frac{\l I_S^n \r}{\la I_S\ra}.
\label{F}
\ee

The instanton equation (\ref{WDR-instanton}) may be brought to dimensionless form 
\be
\frac{E_C\lambda}{\hbar}\,\frac{\la F(\lambda {\dot \phi})\ra_E}{ \la(\lambda\dot \phi)^2 \ra_E}= 
{\cal P} \equiv \frac{eG_L}{4CJ_S},
\label{instanton}
\ee
where $E_C=e^2/2 C$ is the capacitor's charging energy, and we define the threshold parameter ${\cal P}$, 
which is the ratio of the JJ energy relaxation rate $G_L/C$ to the mesoscopic system's excitation rate $J_S/e$.
If relaxation is weak, ${\cal P} \ll 1$, then Eqs.\ (\ref{F}) and (\ref{instanton}) may be truncated at the first (Gaussian) term,
giving the escape rate (\ref{WDR-instanton}) as
\be
\log \Gamma_{\rm G} = -\frac{\Delta U}{T_{\rm eff}},
\label{gauss}
\ee
where $T_{\rm eff} = \l I_S^2\r/ 2G_L$ is an effective noise temperature, 
so the Arrhenius form is recovered in agreement with \cite{SJthreshold1}.  
However, if the relaxation rate dominates the excitation rate so ${\cal P} \gg 1$, 
then we enter the {\it threshold regime} where the solution to (\ref{instanton}) 
is non-perturbative in the cumulant expansion.  

The physical reason for this is that while the Gaussian contribution to the 
system noise cannot compensate for the fast energy relaxation, rare current events (where many electrons are 
sequentially transmitted in a short amount of time) may be able to  
excite the JJ strongly enough to overcome damping. The power 
of these current kicks is proportional to the velocity $\dot\phi$, and 
therefore creates positive feedback for rare-event activation in the weak damping regime, because the velocity 
continues to grows as the particle gradually ascends the potential well to the escape point. 

By introducing the JJ quality factor ${\cal Q}= \omega_{\rm pl} C/G_L>1$
and the separation of time scales parameter ${\cal R}=eV_S/\hbar\omega_{\rm pl}>1$
we reformulate the condition for the threshold regime as 
${\cal P}=(1/{\cal QR})(e^2/4\hbar G_S)>1$, therefore $G_S<e^2/\hbar$. 
This implies that the system is a tunnel junction,
which is known to create Poissonian noise (however, see the discussion below).
The circuit back action can be neglected if $eV_S>eV=e\Phi_0\dot\phi\sim \sqrt{E_CE}$.
We will show below that $E\sim E_C$ at the threshold, so 
$eV_S/E_C={\cal QR}(\hbar G_L/e^2)>1$. Together with ${\cal P}>1$
this gives the overscreening condition $G_L>G_S$.    

{\it Poissonian noise.}---
In order to illustrate the threshold behavior, we consider a simple harmonic potential 
$U(\phi) = (1/2)m \omega_{\rm pl}^2 \phi^2$ with a sharp cut-off at $\phi = \phi_0$. 
In this simplified model 
$\phi(t) = \sqrt{2 E/E_J} \cos(\omega_{\rm pl} t)$. The averages in 
Eq.~(\ref{instanton}) over the periodic orbit at constant energy may 
now be done exactly.  The Poissonian noise generator is $F(z) = e^z -z -1$.
Using the integral representation of the modified Bessel function of order 0, 
$I_0(z) = (1/2\pi) \int_0^{2 \pi} d \theta \exp(-z \sin \theta)$, 
we find Eq.~(\ref{instanton}) simplifies to
\be
 \frac{I_0(z) -1}{z} = 2{\cal P}\sqrt{\frac{E}{E_C}},\quad  z\equiv\lambda\sqrt{\frac{2E}{m}}.
\label{bessel}
\ee
This equation is numerically solved for $\lambda$ and integrated over the energy $E$. The result 
for the logarithm of the escape rate is plotted in Fig.\ \ref{circuit} 
as a function of $\Delta U$ for 
different values of $\cal P$.  

For ${\cal P} \ll 1$, the Bessel function may be expanded as $I_0(z) = 1 + z^2/4 + \ldots$, 
which gives $\lambda = 2\hbar {\cal P}/E_C$.  
When $\lambda$ is substituted into (\ref{WDR-instanton}), the result (\ref{gauss}) is recovered, 
because for Poissonian noise $\l I^2_S \r =G_SV_S$.
In the opposite limit, ${\cal P} \gg 1$, the right hand side of Eq.\ (\ref{bessel}) is large, and 
we utilize the asymptotic form of the Bessel function, $I_0(z) \rightarrow e^z/\sqrt{2 \pi z}$ for $z$ large.
Thus the variable $z$ must be only a logarithmically growing solution of Eq.\ (\ref{bessel}):
$z=L_{\rm H}(E)\equiv\log(2{\cal P}\sqrt{2\pi E/E_C}\,)$, up to the double-logarithmic correction.
This gives $\lambda =L_{\rm H}(E)\sqrt{m/2E}$, so the escape rate (\ref{WDR-instanton}) of 
activation due to Poissonian noise may be approximated by 
\be
\log \Gamma_{\rm P} = - \sqrt{\frac{\Delta U}{E_C}} 
\log\left(2{\cal P} \sqrt{\frac{2\pi\Delta U}{E_C}} \right).
\label{threshold-rate}
\ee 
%This behavior clearly differs from the Arrhenius form (\ref{gauss}). 
The dominant square-root behavior is clearly seen in Fig.\ \ref{circuit}, and is a sign of the break-down of 
Gaussian noise activation. 

For the escape to be achievable, we require that $\Delta U$ is not
much larger than $E_C$. More rigorously, for strong bias
the action scales as $\sqrt{\Delta U/E_C}\sim(1-{\cal J})^{3/4}\sqrt{E_J/E_C}$.  
The total number of states in the quantum well, $N_{\rm tot}\sim\Delta U/\hbar\omega_{\rm pl}$,
also scales down as $N_{\rm tot}\sim(1-{\cal J})^{5/4}\sqrt{E_J/E_C}$, but 
must be large in the semiclassical limit. Therefore, we estimate 
$-\log\Gamma_{\rm P}\sim N_{\rm tot} \log{\cal P}/\sqrt{1-{\cal J}_{\rm st}}$.  In Refs.\ \cite{Pekola2,Saclay} this number varies from 10 to 17. This indicates that it may be hard but nevertheless feasible to observe the threshold behavior.  

{\em Anharmonic correction.---}
 Here we show that a realistic potential $U(\phi)$ leads to only a small anharmonic correction to the escape rate plotted in Fig.\ \ref{circuit}. 
In the case of a general potential, 
the instanton line Eq.~(\ref{instanton}) may be found by noting that for Poissonian noise, the integral 
in $\la F(\lambda {\dot \phi})\ra_E$ in the threshold regime is exponentially dominated by 
the largest value of ${\dot \phi}$.   Energy conservation, $(m/2){\dot \phi}^2 + U(\phi) = E$, 
indicates that this will be near the bottom of the potential, where the potential is approximately 
harmonic with frequency $\omega_{\rm pl}$.  Therefore, the previous Bessel function result will hold 
to excellent approximation.  The other average, $\la {\dot \phi}^2\ra_E$, is generalized by noting 
that by using $dt = d\phi/|{\dot \phi}|$ and conservation of energy, the equation 
$\partial_E [T_{\rm p}(E) \la {\dot \phi}^2\ra_E] = T_{\rm p}(E)/m$ holds.  
This equation may be integrated to find 
$\la {\dot \phi}^2\ra_E = 2\pi \hbar N(E)/m T_{\rm p}(E)$,
where $N(E)=(1/2 \pi \hbar) \int_0^E dE' T_{\rm p}(E')$ is the number of quantum states 
in the cavity below the energy $E$.  
%We note that quantum mechanics tells us that $N(E) \ge E/\hbar \omega_{pl}$, where equality 
%recovers the previous results. 

Combining results for $\la F\ra_E$ and $\la \dot \phi^2\ra_E$ and 
going to the asymptotic threshold regime, we conclude that the instanton solution takes 
the previously found form $\lambda =L(E)\sqrt{m/2E}$ with the logarithm replaced by
$L(E)=L_H(E) - \log[A(E)]$, where the function 
\be
A(E) = \frac{T_{\rm p}E}{2\pi\hbar N}=E\,\partial_E\log N
\ee 
characterizes the anharmonicity. It takes the value $A=1$ at the 
bottom of the potential well and diverges logarithmically at the top of the barrier
as $A\sim \log [\Delta U/(\Delta U-E)]$. Thus this factor may compensate the large parameter 
${\cal P}$ in close vicinity of the barrier top, $\Delta U-E\sim \Delta Ue^{-\cal P}$,
so the Gaussian noise there dominates. However, the overall anharmonic contribution to 
the large logarithm $L(E)$ is relatively small and can be neglected.  

{\it Stabilization effects.}---
In what follows we wish to illustrate an important fact: statistics of rare 
current fluctuations carries complementary information about a Markovian process 
which is not contained in any finite current cumulant. As a first example, we consider the shot noise from a quantum point contact (QPC), which is known to be a binomial process.  The noise Fano-factors are generated by $F(z) = (1/{\cal T}) \log[1+{\cal T}(e^z-1)] - z$, where ${\cal T}$ 
is the transmission of the QPC. The cumulants are obtained by the expansion (\ref{F})
at $z=0$. Therefore, in the tunneling regime, ${\cal T}\ll 1$, the logarithm can be
expanded to lowest order in ${\cal T}$ so the noise of the QPC is Poissonian, 
$F(z) = e^z-z-1$, as in the example considered above. However, our results indicate 
that the Poissonian process can always provide a strong enough current
fluctuation for the JJ to escapes from the metastable state with some small probability.  
This is not the case for a QPC in the tunneling regime,
%.  For any finite value of ${\cal T}$, the extreme value statistics (and thus the escape rate of the JJ) can be quite different 
as we demonstrate below.

Indeed,  the rare current events of the QPC are determined by the asymptotic of $F(z)$ 
at large $z$:
\be
F(z)+z = \frac{1}{{\cal T}}\begin{cases} z + \log {\cal T}, & z\to+\infty, \\ \log (1 - {\cal T}), & z\to-\infty. 
\end{cases}
\label{Binomial}
\ee
This result implies that with a small probability ${\cal T}^M$ the current acquires its maximum value
$J_{\rm max} = J_S/{\cal T}=e^2V_S/\pi\hbar$, so that all $M=eV_S\Delta t/\pi\hbar$ electrons arriving at 
the QPC during time interval $\Delta t$ are transmitted. Similarly, with the probability
$(1-{\cal T})^M$ all the electrons are reflected giving zero current \cite{footnote}. If $\Delta U$
exceeds some critical value, the maximum or minimum current fluctuation
creates insufficient bias for the JJ to escape from the supercurrent state \cite{Nazarov}. 
We named this phenomenon the {\em Pauli stabilization effect} because it originates from
the Pauli principle of electron occupation \cite{ourPRB}.

To estimate the value of the bias, ${\cal J}_{\rm st}$, and escape rate, $\Gamma_{\rm st}$, 
at the stabilization point, we average $F$ in (\ref{Binomial}) over the energy-conserving 
trajectories, which yields 
$\la F\ra_E = (1/{\cal T}) \left\{ \lambda \Delta \phi/T_{\rm p} + (1/2)\log [{\cal T} (1-{\cal T})]\right\}$, where 
$\Delta \phi$ is the distance between the two turning points at energy $E$.  
The denominator of (\ref{instanton}) is the same as before, so the instanton line is given by 
\be
\lambda_{\rm st}(E) = \frac{aT_{\rm p}}{\Delta \phi - bN},
\label{lambda}
\ee  
with the coefficients $a=-(1/2)\log[{\cal T} (1-{\cal T})]$ and $b=16\pi {\cal P} {\cal T}$.
Integration of this instanton line leads to singular behavior when $\Delta \phi = 
16\pi{\cal P} {\cal T}N$.  Evaluated at $E=\Delta U$ this equality determines
the stabilization point ${\cal J}_{\rm st}$ below which the rate $\Gamma$ vanishes. 
We find this point by assuming the strong bias limit $1 - {\cal J} \ll 1$ in (\ref{potential})
giving a cubic potential:
$U/E_J=(1-{\cal J})\phi-\phi^3/6$. Skipping a number of steps, we present the result:
\be
 (1 - {\cal J}_{\rm st})^{3/4}= \frac{5\sqrt{E_C/E_J}}{ 32\cdot 2^{1/4} {\cal P} {\cal T}}.
 \label{stab-bias}
\ee

Next, we note that at the critical point ${\cal J}={\cal J}_{\rm st}$ the integral (\ref{WDR-instanton})
with $\lambda=\lambda_{\rm st}(E)$ from Eq.\ (\ref{lambda}) is convergent because $\lambda_{\rm st}(E)$ 
has only has an inverse square-root divergence at $E=\Delta U$. Therefore, $\log\Gamma_{\rm st}$ can be estimated
by dropping the least singular term, $b N$, in the denominator of $\lambda_{\rm st}(E)$ and replacing the potential
with the harmonic approximation $U=(1/2)m\omega_{\rm pl}^2\phi^2$. Straightforward evaluation
then gives $\log\Gamma_{\rm st}\sim \sqrt{\Delta U/E_C}\log[{\cal T} (1-{\cal T})]$. At strong bias 
$\Delta U\sim(1-{\cal J_{\rm st}})^{3/2}E_J$, and using the result (\ref{stab-bias}) we find the escape rate 
right before stabilization:
\be
\log\Gamma_{\rm st}\sim \frac{\log[{\cal T} (1-{\cal T})]}{{\cal P}{\cal T}}. 
\label{stab-rate}
\ee 
We estimate (\ref{stab-rate})  
in terms of the quality factor ${\cal Q}>1$ and the separation of time scales parameter ${\cal R}>1$ as 
$\log\Gamma_{\rm st}\sim {\cal QR}\log[{\cal T} (1-{\cal T})]$. 
Alternatively, using the total number of states $N_{\rm tot}>1$ in the
quantum well, and Eq.\ (\ref{stab-bias}), we find that $\log\Gamma_{\rm
st}\sim N_{\rm tot} \log[{\cal T} (1-{\cal T})]/\sqrt{1-{\cal J}_{\rm st}}$,
which agrees with the previous estimate for Poissonian noise.  
To compare with the Poissonian case, 
the escape rate for a QPC with the transmission ${\cal T}=0.06$ is plotted in 
Fig.\ \ref{circuit} in the case of the simple harmonic potential with a sharp cut-off.
The sharp potential leads to a logarithmic divergence at the stabilization 
point $\Delta U/E_C = 6.86$, in contrast with the rate discontinuity 
discussed above for a smooth potential.

Our second example is the telegraph process: The system current switches randomly from the value $I_1$
to the value $I_2$ and back with the rate $\gamma_1$ and $\gamma_2$, respectively. We have found 
in Ref.\ \cite{SJthreshold2} that the cumulant generator of the telegraph process is given by 
the formula
\be
{\cal H}_{S}(z)=\bar Iz-\bar\gamma+\sqrt{(\Delta Iz-\Delta\gamma)^2/4+\gamma_1\gamma_2},
\label{telegraph}
\ee
where $\Delta I=I_2-I_1$, $\Delta \gamma=\gamma_2-\gamma_1$, $\bar I=(I_1+I_2)/2$, and 
$\bar \gamma=(\gamma_1+\gamma_2)/2$.
In the case of slow switching, $\gamma_1,\gamma_2\ll I_1,I_2$,
the noise becomes super-Poissonian with the current distributed between $I_1$ and $I_2$.
The sharp cut-off of the distribution function at these values results in the stabilization effect.
The asymptotic form ${\cal H}_{S}=\bar I z - \bar \gamma +|\Delta I z -\Delta \gamma| /2$ at $|z|\to\infty$ 
leads to the result (\ref{lambda}) with the new coefficients $a=\bar\gamma/\Delta I$ and $b=8 \pi G_L E_C/e\Delta I$.
Therefore, at the stabilization point the results (\ref{stab-bias}) and (\ref{stab-rate}) hold 
after the replacement ${\cal P}{\cal T}\to G_L E_C/2 e \Delta I$ and 
$\log[{\cal T} (1-{\cal T})]\to -2\bar\gamma/\Delta I$.
With the new separation of time scales requirement, ${\cal R}'=\bar\gamma/\omega_{\rm pl}>1$, 
the action at the stabilization point can be estimated as $-\log\Gamma_{\rm st}\sim {\cal QR}'$,
so the telegraph stabilization effect should also be observable.

%We have demonstrated that it is possible to use a single JJ to access the threshold regime for detecting full counting statistics.  While the experimental requirements are demanding, we found that overscreening the junction enhances the energy relaxation rate, making it comparable to the excitation rate of the junction.   This suppresses the contribution of the Gaussian fluctuations while enhancing the effect of the current tails.

We acknowledge the support of the Swiss NSF and the University of Rochester.

\bibliographystyle{apsrev}

\begin{thebibliography}{99}
\bibitem{footnote} The Binomial process is invariant under the transformation
${\cal T}\to 1-{\cal T}$ and $z\to -z$, therefore weak backscattering is equivalent to weak tunneling.
\end{thebibliography}

\end{document}